\documentclass{WileyMSP-template}
\usepackage{amsfonts}
\usepackage{mathrsfs}
\usepackage{amsmath}
\usepackage{color}
\usepackage{graphicx}
\usepackage{bm}
\usepackage{amssymb}
\usepackage{xspace}
\usepackage{epstopdf}
\usepackage{dcolumn}
\usepackage{multirow}
\usepackage{wrapfig}
\usepackage{cite}
\usepackage{ragged2e}
\usepackage{makecell}

\begin{document}

\pagestyle{fancy}
\rhead{\includegraphics[width=2.5cm]{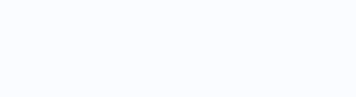}}

\title{Real topological phonons in 3D carbon allotropes}

\maketitle

\author{Xiaotian Wang$^\dagger$,}
\author{Jingbo Bai$^\dagger$,}
\author{Jianhua Wang$^\dagger$,}
\author{Zhenxiang Cheng$^*$,}
\author{Shifeng Qian$^*$,}
\author{Wenhong Wang,}
\author{Gang Zhang,}
\author{Zhi-Ming Yu$^*$,}
\author{Yugui Yao}\\
($^\dagger$These authors contributed equally to this work.)

\dedication{ }

\begin{affiliations}
X. Wang, J. Bai\\
School of Physical Science and Technology, Southwest University, Chongqing 400715, China\\
J. Wang, W. Wang\\
School of Material Science and Engineering, Tiangong University, Tianjin 300387, China\\
G. Zhang\\
Yangtze Delta Region Academy of Beijing Institute of Technology, Jiaxing 314000, China\\
S. Qian\\
Anhui Province Key Laboratory of Optoelectric Materials Science and Technology, Department of Physics, Anhui Normal University, Anhui, Wuhu 241000, China\\
$^*$Email: qiansf@ahnu.edu.cn\\
X. Wang, Z. Cheng\\
Institute for Superconducting and Electronic Materials (ISEM), University of Wollongong, Wollongong 2500, Australia\\
$^*$Email: cheng@uow.edu.au\\
Z. Yu, Y. Yao\\
Key Lab of advanced optoelectronic quantum architecture and measurement (MOE), Beijing Key Lab of Nanophotonics $\&$ Ultrafine Optoelectronic Systems, and School of Physics, Beijing Institute of Technology, Beijing 100081, China\\
$^*$Email: zhiming$\_$yu@bit.edu.cn\\
\end{affiliations}



\begin{abstract}
There has been a significant focus on real topological systems that enjoy space-time inversion symmetry ($\mathcal{P} \mathcal{T}$) and lack spin-orbit coupling. 
While the theoretical classification of the real topology has been established, more progress has yet to be made in the materials realization of such real topological systems in  three dimensions (3D).
To address this crucial issue, by selecting the carbon-based material candidates as targets, we perform high-throughput computing to inspect the real topology in the  phonon spectrums of the  3D carbon allotropes in the Samara Carbon Allotrope Database (SACADA). Among 1192 kinds of 3D carbon allotropes, we find 65 real topological systems with a phononic real Chern insulating (PRCI) state, 2 real topological systems with a phononic real nodal line (PRNL) state, 10 real topological systems with a phononic real Dirac point (PRDP) state, and 8 real topological systems with a phononic real triple-point pair (PRTPP) state.
This  extremely expands the material candidates with  real topology, especially for the  gapless topological phonons.
We exhibit the PRCI, PRNL, PRTPP, and PRDP states of 27-SG. 166-pcu-h, 1081-SG. 194-4$^2$T13-CA, 52-SG. 141-gis, and 132-SG. 191-3,4T157 as illustrative examples, and explore the second-order boundary mode, i.e., phononic hinge mode.
Among the four examples, the materials 1081-SG. 194-4$^2$T13-CA and  52-SG. 141-gis are so ideal that  the PRNL and  PRTPP  in them are well separated from other bands, and the phononic hinge mode can be clearly observed.
This study aims to broaden the understanding of 3D topological phonons, and emphasizes the potential of 3D carbon allotropes as a valuable framework for exploring the fascinating physics related to phononic hinge modes and phononic real topology.

\end{abstract}

\justifying

\section{Introduction}
There has been significant interest in condensed-matter research due to the unique topological states of matter and their material realization~\cite{add1,add2,add3,add4,add5}.
Unlike the conventional states with vanishing geometric quantities, the topological states in solid exhibit nontrivial  geometric quantities and are characterized by  global topological quantities.
One of the  most famous  topological quantities is the Chern number, which is not only important in mathematics but also  is directly connected to the Hall conductivity of the systems with finite Chern number, i.e., the Chern insulator.
Besides, one can use the Chern number to classify the nodal points in three dimensions, such as Weyl point,  spin-1 fermion, and many unconventional  points.
For the systems with vanishing Chern number, they can be classified by the higher-order topological quantities.
The study has  been extended to real topology, which is a special kind of higher-order topology, that appears in materials with  space-time inversion symmetry ($\mathcal{P} \mathcal{T}$) and lack spin-orbit coupling.
The real topology  originates in the mathematical structure of the real-valued vector bundles associated with the real-valued wave functions, belonging to  the Stiefel-Whitney class~\cite{add13}.
Similar to the Chern number, the real Chern number ($\nu_R$) is defined in a 2D closed surface. Thus, there exist many kinds of real topological systems in 3D~\cite{add7,add8,add19,add19a}, including  real Chern insulator (RCI), real nodal line (RNL), real Dirac point (RDP),  and real  triple-point pair (RTPP) systems.

Two equivalently important tasks in the field are classifying all possible topological states that are protected by crystal group symmetries, and identifying the material candidates hosting  nontrivial topology.
The classification of the topological states, including first-order and higher-order topologies, is established.
Furthermore,  several high-throughput computation works have been performed to investigate the  electronic band structure of  all stable non-magnetic and magnetic solid materials, resulting in thousands of possible topological electronic materials~\cite{add7a,add7b,add7c,add7d,add7e,add7ee,add7ef,add7eg,add7eh}. Recently, Xu $et$ $al.$~\cite{add7f} studied the topological states in the phonon spectrum of more than 10,000 materials and found nontrivial phononic bands in more than half of the studied materials.


However, most works on the material realization of real topology   thus far are case-by-case type studies, focusing on the electronic and artificial systems~\cite{add8,add12,add14,add15,add15a,add16,add17,add18,add19,add20,add21,add21b}, including RCIs, RNL semimetals, RDP semimetals, and RTPP semimetals.
For example, Zeng $et$ $al.$~\cite{add16} proposed bulk $\gamma$-graphyne as a realistic candidate for 3D RCIs.  Chen $et$ $al.$~\cite{add12} presented the experimentally synthesized 3D graphdiyne as the first example of RNL semimetal, using first-principles calculations and theoretical analysis. Qian $et$ $al.$~\cite{add18} challenged a common belief that RDP is unstable and is commonly regarded as the critical point of a RNL. They also reported that 3D twisted bilayer graphene is an example of RDP semimetals. Lenggenhager $et$ $al.$~\cite{add19} proposed the RTPP state, a triple-point pair characterized by a nontrivial $\nu_R$. Additionally, they stated that 3D Sc$_3$AlC is a potential candidate for RTPP semimetals.

An emerging trend in topological materials research is to shift the focus from electronic \cite{add10,add22,add23,add24,add25,add26} to phononic systems, with the goal of investigating topological phonons and related phononic applications~\cite{add27,add28,add29,add30,add31,add32}, as  the ideal real topology in electronic materials is very limited due to the strong Fermi level constraint. Moreover, the SOC effect is generally difficult to ignore in electronic systems. Nevertheless, the phonons, which are intrinsically in the absence of the SOC  and the limitation imposed by the Fermi level, offer a robust platform for conducting $\mathcal{P} \mathcal{T}$-symmetric searches to identify material samples that possess real topology across the whole frequencies of the phonon spectrum.
Most current  research focuses on the investigation of gapless topological phonons in solids, particularly the identification of gap-closing points, lines, or surfaces based on the crossings of phononic bands~\cite{add33,add34,add35}.
Moreover, real gapped topological phonon states, i.e., the phononic real Chern insulating (PRCI) states, in 2D materials~\cite{add36,add37,add38,add39} have received significant attention from researchers, and there is a shortage of studies on the PRCI states in 3D materials.
Regrettably, the real gapless topological phonon states, which possess a $\mathbb{Z}_2$ monopole charge and are classified under the $\nu_R$, are rarely reported by other researchers in 3D materials.

In order to establish the feasibility of achieving real gapped and gapless topological phonons in 3D materials, this study focuses on a comprehensive set of 1192 carbon allotropes sourced from the SACADA~\cite{add21a,add40}. By a calculation of the phonon spectrum of these 3D carbon allotropes, we identify the appearance of the real gapped topological phonon state, referred to as the PRCI state, as well as three real gapless topological phononic states, namely the phononic real nodal line (PRNL), phononic real triple-point pair (PRTPP), and phononic real Dirac point (PRDP) states, in 3D materials. Specifically, our findings indicate that among the 1192 candidates, 65 possess a PRCI state, 2 possess a PRNL state, 10 possess a PRDP state, and 8 possess a PRTPP state, respectively. The nontrivial real topology in bulk manifests as topological boundary modes located at a pair of hinges of a bulk sample. In this work, second-order phononic hinge modes are validated by performing phononic tight-binding computations on a 1D tube geometry using 27-SG. 166-pcu-h, 1081-SG. 194-4$^2$T13-CA, and 52-SG. 141-gis as sample cases. It is important to highlight that this is the first attempt to forecast phononic hinge modes in 3D systems. Furthermore, the material samples 1081-SG. 194-4$^2$T13-CA, and 52-SG. 141-gis represent the first examples with ideal PRNL and PRTPP states, respectively. This study expands the scope of 3D topological phonons that can possess a nontrivial $\nu_R$ protected by the $\mathcal{P} \mathcal{T}$ symmetry. It highlights the significance of 3D carbon allotropes as a good platform for investigating the intriguing physics associated with phononic hinge modes and phononic real topology.

\section{Results and discussion}

\subsection{Classifications of real topological phonons}

\begin{figure}[h]
\centering
  \includegraphics[height=12cm]{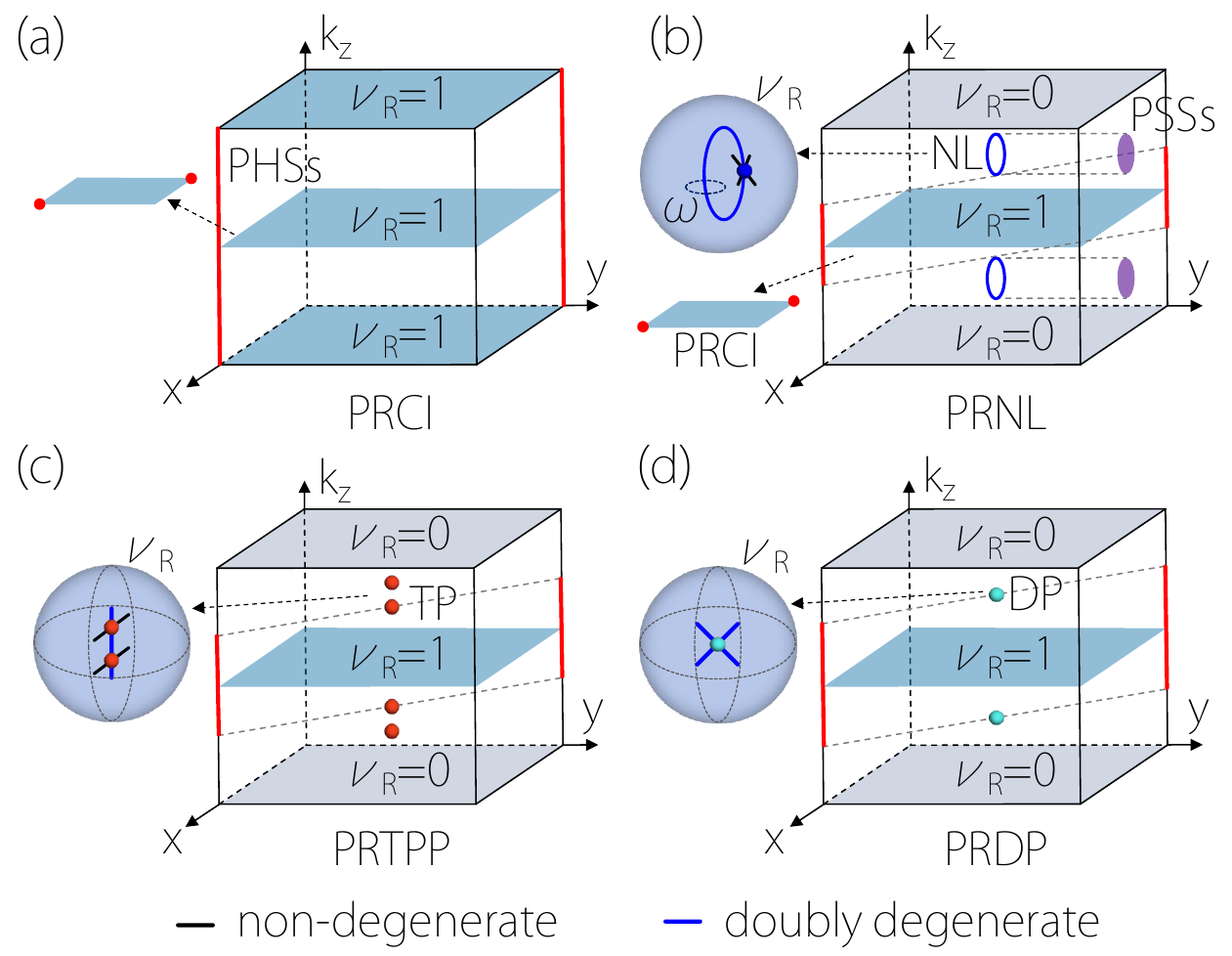}
  \caption{Four cases of real topological phonons and related second-order phononic hinge states (PHSs) as highlighted by red lines. (a), (b), (c), and (d) are the schematic figures of PRCI, PRNL, PRTPP, and PRDP states, respectively.}
  \label{fig1}
\end{figure}
$\textbf{Figure~\ref{fig1}}$a shows the schematic figure of a PRCI state in 3D. The real topology of 3D PRCI can be determined by the $\nu_R$  for a generic  2D plane along one direction, as the system is insulting. We select $k_z=0$  plane as an example, the $\nu_R$ of it  can be readily extracted from the parity eigenvalues at the time-reversal invariant momentum (TRIM) points on $k_z=0$ plane, with~\cite{add7,add8}
\begin{equation}
(-1)^{\nu_R}=\prod_i(-1)^{\left\lfloor\left(n_{-}^{\Gamma_i} / 2\right)\right\rfloor},
\end{equation}
where $\lfloor\cdots\rfloor$ is the floor function, and $n_{-}^{\Gamma_i}$ is the number of occupied bands with negative $\mathcal{P} \mathcal{T}$ eigenvalue at TRIM point ${\Gamma_i}$. If one finds that the $\nu_R$ is nontrivial ($\nu_R$ = 1) for  $k_z=0$ plane, which confirms the PRCI nature. Furthermore, the nontrivial $\nu_R$ results in second-order phononic hinge states (PHSs) (red lines) along the entire Brillouin zone (BZ) of the $k_z$ direction. Actually, each 2D $k_z$ plane can be viewed as a 2D PRCI, which has two zero modes (red dots) at a pair of $\mathcal{P} \mathcal{T}$-related corners.

$\textbf{Figure~\ref{fig1}}$b shows the schematic figure of a PRNL state in 3D. Each 2D $k_z$ plane between the nodal lines is a 2D PRCI with $\nu_R$ = 1 and with a pair of $\mathcal{P} \mathcal{T}$-related corners. Moreover, each nodal line carries a nontrivial  $\nu_R$ = 1 defined on a small sphere enclosing one of the nodal lines. Besides the nontrivial $\nu_R$, each nodal line carries a nontrivial first Stiefel-Whitney number $w$, corresponding to the quantized $\pi$ Berry phase, with

\begin{equation}
w=\frac{1}{\pi} \oint_C \operatorname{Tr} \mathcal{A}(\boldsymbol{k}) \cdot d \boldsymbol{k} \quad \bmod 2,
\end{equation}
where $\mathcal{A}$ is the Berry connection for the occupied phononic bands and $C$ is a closed path encircling the line. Hence, the PRNL states exhibit two mathematical invariants $w$ and the $\nu_R$, and the two topological charges result in boundary states of different dimensions, i.e., phononic surface states (PSSs) and PHSs, respectively.

Note that a PRNL can shrink to a PRDP or a PRTPP and stabilize in materials with additional crystalline symmetries. $\textbf{Figure~\ref{fig1}}$c and $\textbf{Figure~\ref{fig1}}$d show the schematic figures of PRDP and PRTPP states in 3D. Each 2D $k_z$ plane between the two TPPs (DPs) is a 2D PRCI with $\nu_R$ = 1 and a pair of $\mathcal{P} \mathcal{T}$-related corners. These corner states constitute the two PHSs for the 3D system. Each TPP (DP) carries a nontrivial $\nu_R$ = 1 defined on a small sphere enclosing one of the TPPs (DPs), as shown in $\textbf{Figure~\ref{fig1}}$c ($\textbf{Figure~\ref{fig1}}$d).

Note that the $\nu_R$ of $k_z=0$ is nontrivial, whereas the $\nu_R$ of $k_z=\pi$ is trivial, as exhibited in the schematic figures of PRNL, PRTTP, and PRDP (see $\textbf{Figure~\ref{fig1}}$b-d). The switch in topology along the $k_z$ direction from $\nu_R$ = 1  to $\nu_R$ = 0 also reflects the existence of PRNL, PRTTP, and PRDP states between the two planes.

\subsection{The flowchart for screening candidates}
 \begin{figure}[h]
\centering
  \includegraphics[height=9cm]{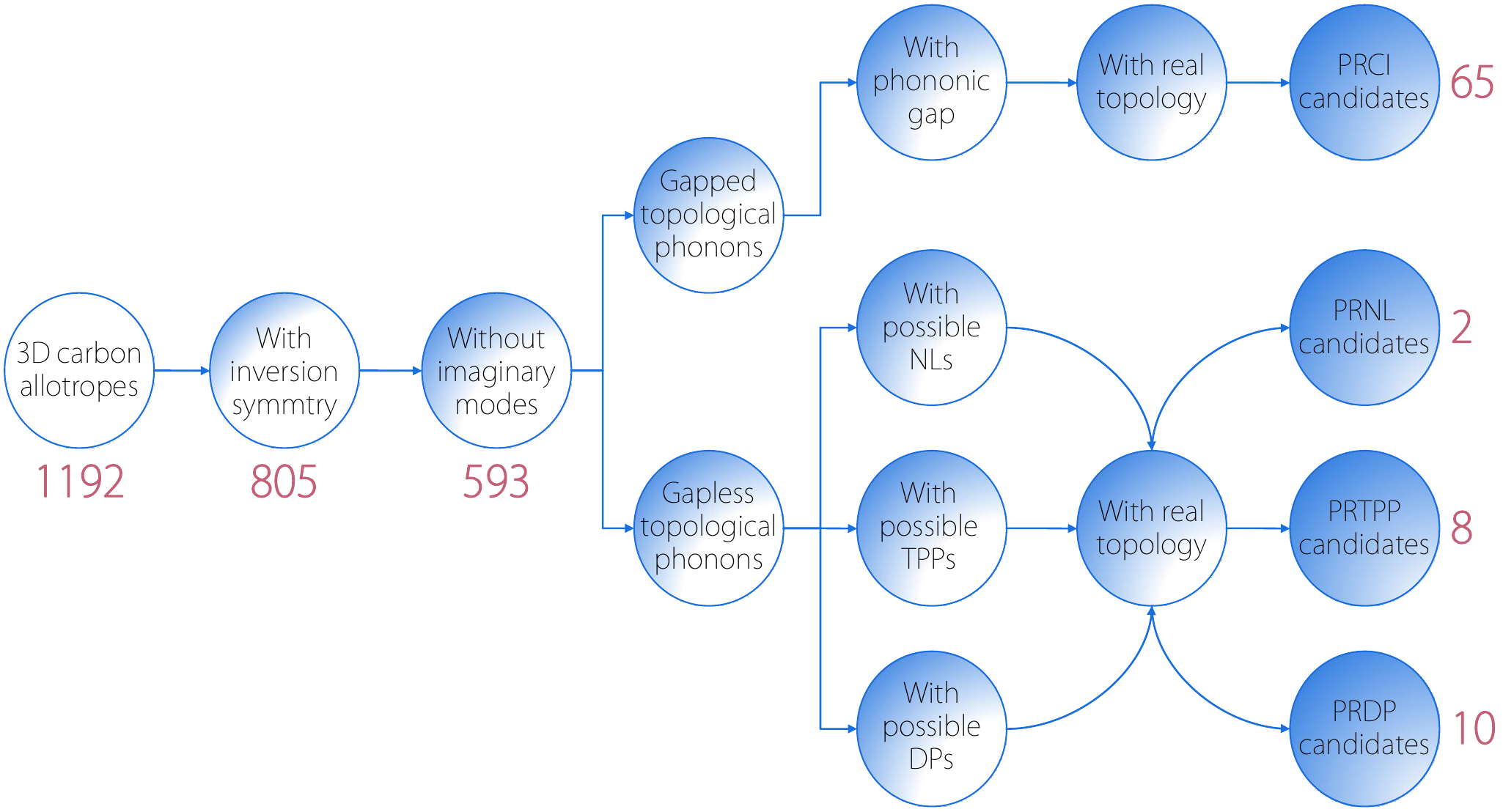}
  \caption{The flowchart for screening 3D carbon allotropes with real topological phonon states, i.e., PRCI, PRNL, PRTPP, PRDP states.}
  \label{fig2}
\end{figure}

Unfortunately, a material realization of the above-mentioned real topological phononic state is rarely touched by other researchers. Our work aims to conduct a material screening to significantly advance the prediction of real topological phonons in 3D carbon allotropes. Over 1000 carbon allotropes have been predicted or synthesized, in addition to the commonly recognized ones like graphite~\cite{add41}, diamond~\cite{add42}, carbon nanotubes~\cite{add43}, graphene~\cite{add44}, and fullerenes~\cite{add45}.

The material screening procedure adheres to the flowchart depicted in $\textbf{Figure~\ref{fig2}}$. The crystallographic data for all carbon allotropes can be obtained from the SACADA database~\cite{add21a,add40}, which offers a full inventory of 1192 3-periodic carbon allotropes. In order to obtain real 3D topological phonons, a total of 805 carbon allotropes possessing $\mathcal{P}$ symmetry are initially chosen. Next, we will determine if the 805 carbon allotropes have any imaginary frequencies by calculating their phonon spectrums. As shown in $\textbf{Figure~\ref{fig2}}$, a selection of 593 3D carbon allotropes that are dynamically stable (without imaginary frequencies) have been chosen for further screening.

Then, the 593 3D carbon allotropes can be used as potential candidates to achieve the gapped and gapless topological phonons. One can identify gaps or the points/lines where phononic bands intersect and close the gap in these 3D carbon allotropes, and examine the real topology for these gapped and gapless topological phonons. As shown in $\textbf{Figure~\ref{fig2}}$, a total of 65 3D carbon allotropes are considered potential candidates with PRCI state since their gaps exhibit a real topology ($\nu_R$ = 1). In addition, by the examination of the presence of potential DP, NL, and TPP, together with their corresponding real topology ($\nu_R$ = 1), it has been demonstrated that 10, 2, and 8 3D carbon allotropes possess PRDP, PRNL, and PRTPP states, which are classified as real gapless topological phononic states.

\subsection{PRCI state in 27-SG. 166-pcu-h}
\begin{figure}[h]
\centering
  \includegraphics[height=9cm]{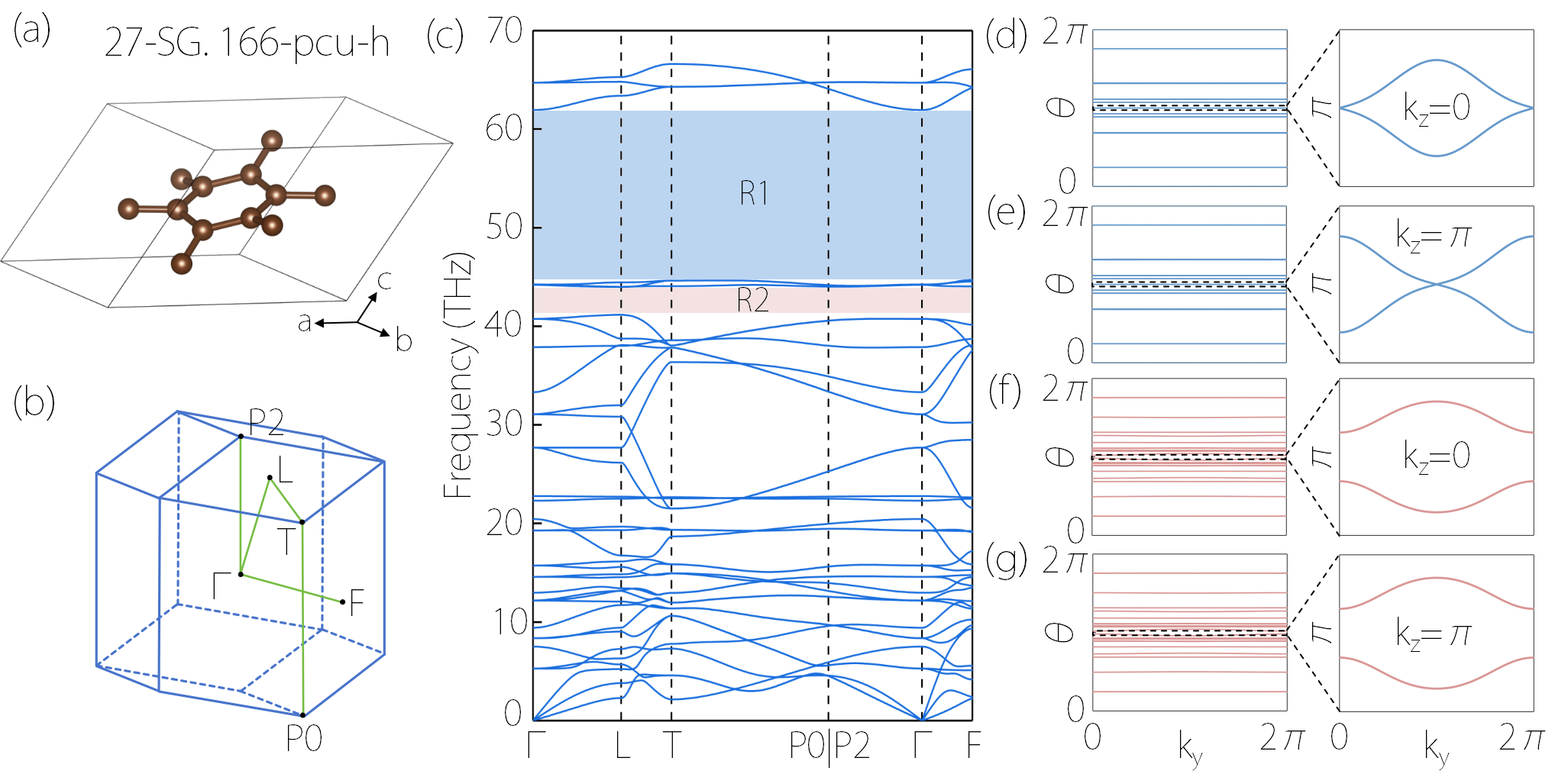}
  \caption{(a) Crystal structure of 27-SG. 166-pcu-h. (b) 3D BZ and selected symmetry paths. (c) Phonon dispersion for 27-SG. 166-pcu-h. Two phononic band gaps (named R1 and R2) are shown. (d)-(g) Wilson loop spectrums calculated on $k_z=0$ and $k_z=\pi$ planes. (d) and (e) for the R1 gap, and (f) and (g) for the R2 gap.}
  \label{fig3}
\end{figure}

\begin{figure}[h]
\centering
  \includegraphics[height=11cm]{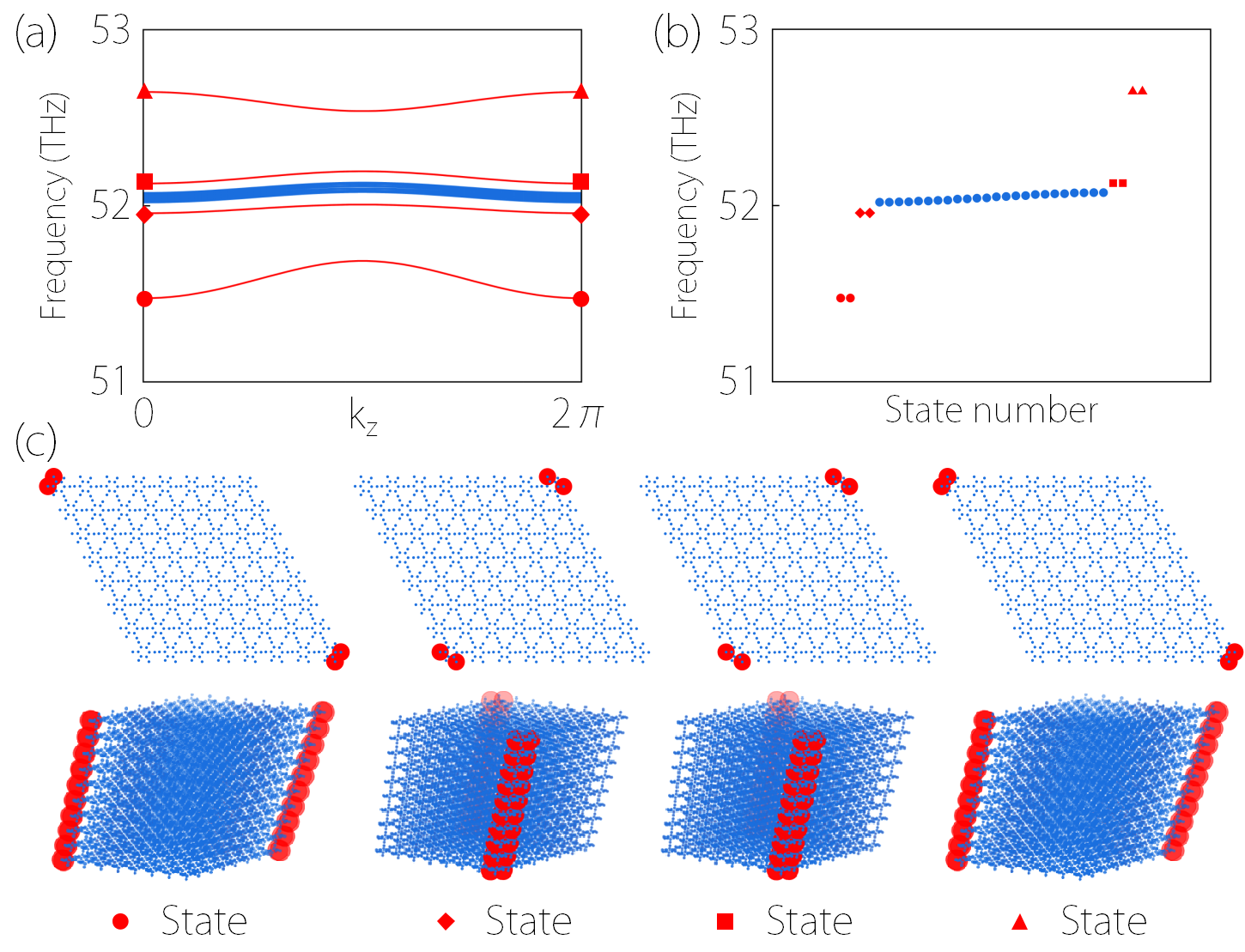}
  \caption{(a) Phonon spectrum (in $k_z$ direction) within the frequency ranges of 51-53 THz for a sample of 27-SG. 166-pcu-h with the tube-like geometry (as exhibited in (c)). Four phononic hinge bands are shown in red. (b) Frequency spectrum spanning the 51-53 THz for four states at $k_z=0$. Distinct shapes denote four doubly degenerate states. (c) The spatial distributions for the four groups of doubly degenerate states at $k_z=0$ under the top and side viewpoints.
  }
  \label{fig4}
\end{figure}

\begin{table}
		\caption{Parity information of the 27-SG. 166-pcu-h at the eight TRIM points.}
		\label{Table1}
\begin{tabular*}{\textwidth}{@{\extracolsep{\fill}} ccccccccc}
  \hline\hline
   & \multicolumn{4}{c}{$k_z$=0}  & \multicolumn{4}{c}{$k_z=\pi$} \\
  \cline{2-5}\cline{6-9}

  Gap R1& $\Gamma$ & $X$ & $M$ & $Y$ & $Z$ & $U$ & $R$ & $T$ \\ \hline
  $n_{+}$ & 15 & 17 & 17 & 17 & 17 & 15 & 17 & 17 \\
  $n_{-}$ & 18 & 16 & 16 & 16 & 16 & 18 & 16 & 16 \\
  $\nu_R$  & \multicolumn{4}{c}{1} & \multicolumn{4}{c}{1}\\
  \cline{2-5}\cline{6-9}
 Gap R2& $\Gamma$ & $X$ & $M$ & $Y$ & $Z$ & $U$ & $R$ & $T$ \\ \hline
  $n_{+}$ & 15 & 15 & 15 & 15 & 15 & 15 & 15 & 15 \\
  $n_{-}$ & 15 & 15 & 15 & 15 & 15 & 15 & 15 & 15 \\
  $\nu_R$  & \multicolumn{4}{c}{0} & \multicolumn{4}{c}{0}\\
  \hline\hline
\end{tabular*}
\end{table}

The presence of a 3D PRCI state was identified in 65 out of 1192 carbon allotropes, as elaborated in the Supporting Information (see $\textbf{Tables S1-S65}$ and $\textbf{Figures S1-S64}$). This section will concentrate on an illustrative example, 27-SG. 166-pcu-h, with the intention of providing a thorough comprehension of the real gapped topological phonons and the associated PHSs. The phonon dispersion of 27-SG. 166-pcu-h is shown in $\textbf{Figure~\ref{fig3}}$c, in which two phononic gaps (denoted as R1 and R2) are obvious. In order to examine the topological character of the R1 and R2 gaps, we assess the $\nu_R$ for a 2D slice of BZ with fixed $k_z$. In light of the system's global band gap, it can be inferred that all 2D $k_z$-slices are adiabatically connected, hence necessitating a shared $\nu_R$. That is, one can select a particular slice, such as $k_z=0$ or $k_z=\pi$ plane, to determine the $\nu_R$.

Here, both the $k_z=0$ and $k_z=\pi$ planes are considered, and each one contains four TRIM points $\Gamma_i$. Based on Equation (1), the $\nu_R$ for the  $k_z=0$ and $k_z=\pi$ planes of R1 and R2 gaps are calculated and exhibited in $\textbf{Table~\ref{Table1}}$. At first glance, one finds that the  $k_z=0$ and $k_z=\pi$ planes of R1 (R2) have the same $\nu_R$, as expected. Furthermore, gap R1 hosts a nontrivial $\nu_R$, i.e., $\nu_R$ = 1 for both planes, reflecting the appearance of a PRCI state. However, gap R2 has $\nu_R$ = 0 for both planes, indicating that it is trivial. Besides the parity information at the eight TRIM points for R1 and R2 gaps, we also examine the $\nu_R$ using the Wilson-loop method. The Wilson loop spectrums on $k_z=0$ and $k_z=\pi$ planes of R1 and R2 gaps are calculated, and the results are shown in $\textbf{Figure~\ref{fig3}}$d-g. In the case of the R1 gap, the Wilson loop spectrums in both planes show one crossing point at $\theta=\pi$ , which suggests that $\nu_R$ = 1. However, in the case of the R2 gap, there is no crossing point at $\theta=\pi$ in the Wilson loop spectrums in both planes. This indicates that the value of $\nu_R$ is 0. These results match with what is obtained from the parity information at the TRIM points in $\textbf{Table~\ref{Table1}}$.

Interestingly, second-order phononic boundary states appear in 3D PRCI candidate 27-SG. 166-pcu-h, thanks to the nontrivial $\nu_R$. In order to determine the PHSs for 27-SG. 166-pcu-h, a phononic tight-binding (TB) model for 1D tube geometry was built based on a sample of 27-SG. 166-pcu-h preserving its $\mathcal{P} \mathcal{T}$ symmetry (see $\textbf{Figure~\ref{fig4}}$c). Following this, the phonon band calculation for the nanotube is performed, and $\textbf{Figure~\ref{fig4}}$a displays an enlarged view of the phonon spectrum (in $k_z$ direction) within the frequency ranges of 51-53 THz. Four phononic bands, denoted by red lines in $\textbf{Figure~\ref{fig4}}$a, are proved to be doubly degenerate (see $\textbf{Figure~\ref{fig4}}$b). In $\textbf{Figure~\ref{fig4}}$c, we depict the spatial distribution of the four doubly degenerate states of $\textbf{Figure~\ref{fig4}}$a at $k_z=0$, and one finds that these four groups of states are four pairs of $\mathcal{P} \mathcal{T}$-related hinges.

\subsection{PRNL state in 1081-SG. 194-4$^2$T13-CA}

\begin{figure}[h]
\centering
  \includegraphics[height=10cm]{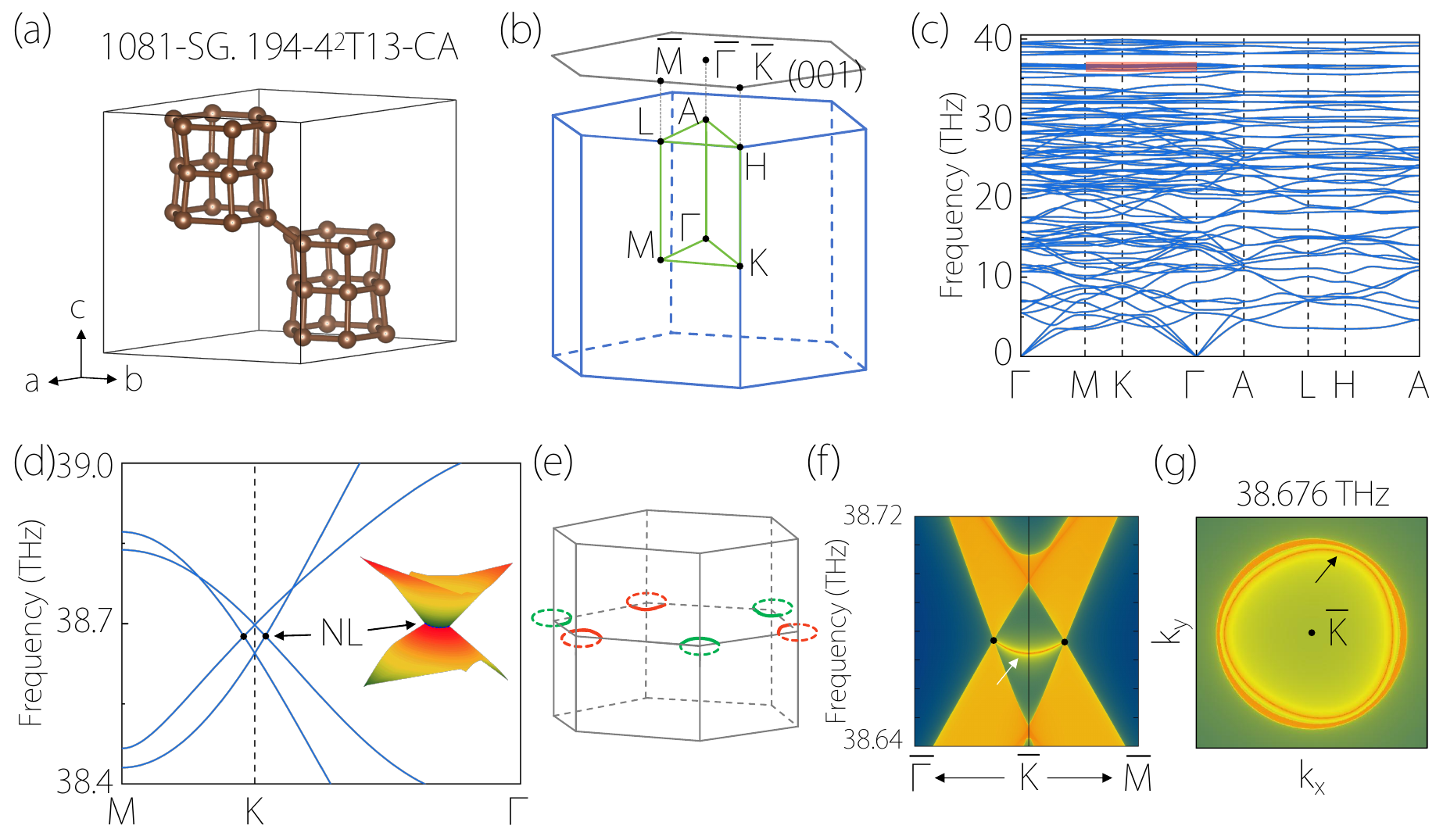}
  \caption{(a) Crystal structure of 1081-SG. 194-4$^2$T13-CA. (b) 3D bulk BZ and the projection of the (001) plane. (c) Calculated phonon spectrum for 1081-SG. 194-4$^2$T13-CA along the symmetry paths $\Gamma$--M--K--$\Gamma$--A--L--H--A. (d) Enlarged phonon spectrum of 1081-SG. 194-4$^2$T13-CA within the frequency ranges of 38.4-39.0 THz. Insert figure shows the 3D plot of the phononic bands around the K-centered closed nodal line. (e) The schematic figure for the $\mathcal{T}$-reversed pair of PRNLs around the K and K' on the $k_z=0$ plane. (f) Projected spectrum on the (001) surface of 1081-SG. 194-4$^2$T13-CA. (g) Constant-frequency slice at 38.676 THz.
  }
  \label{fig5}
\end{figure}

\begin{figure}[h]
\centering
  \includegraphics[height=9.5cm]{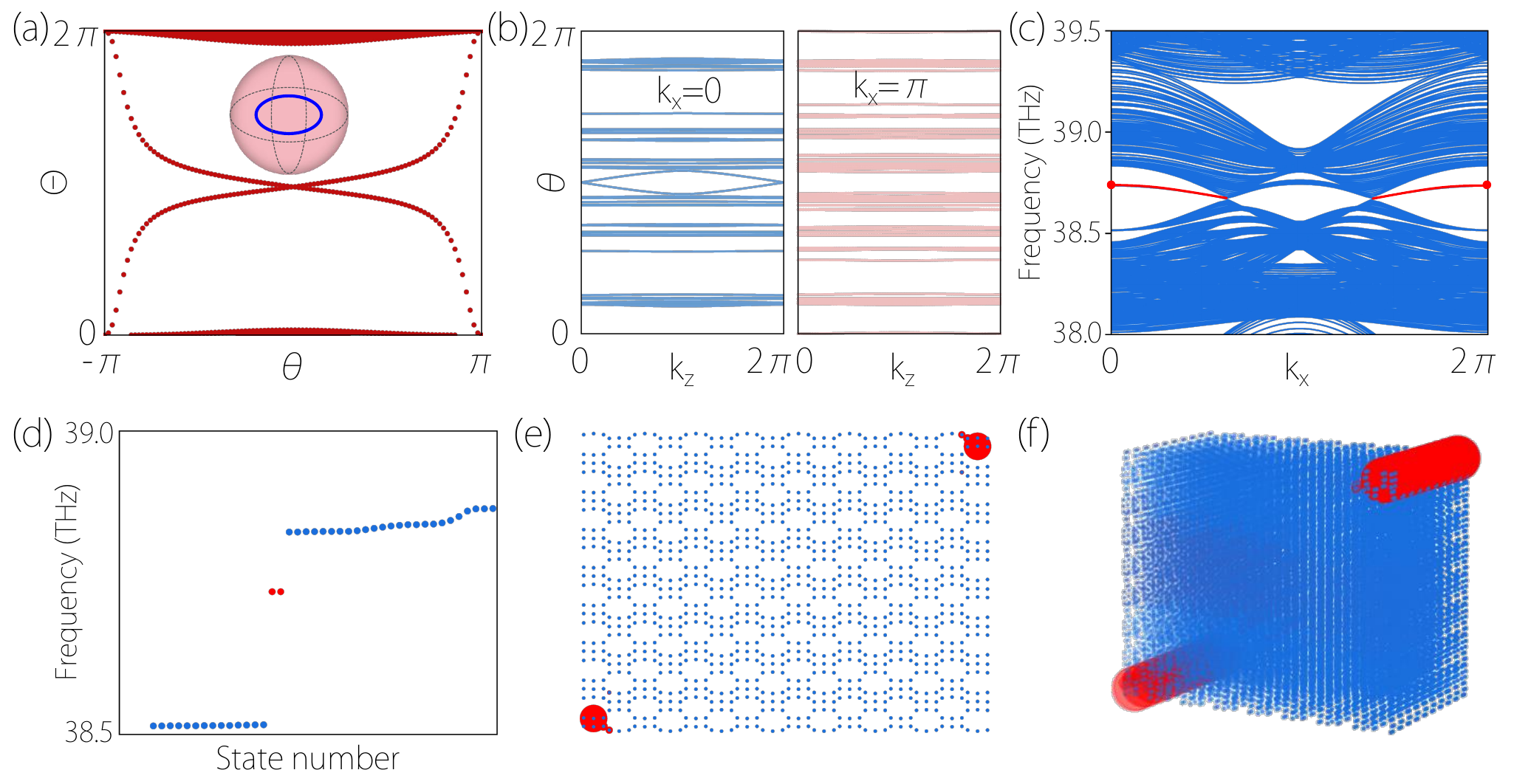}
  \caption{(a) The Wilson loop on a sphere enclosing one of the nodal lines of 1081-SG. 194-4$^2$T13-CA. (b)  Wilson loop spectrums calculated on $k_x=0$ and $k_x=\pi$ planes. (c) Phonon spectrum (in $k_x$ direction) within the frequency ranges of 38-39.5 THz for a sample of 1081-SG. 194-4$^2$T13-CA with the tube-like geometry (as exhibited in (f)). One phononic hinge band is shown in red. (d) Frequency spectrum spanning  for a doubly degenerate state (red dot) at $k_x=0$. (e) and (f) The spatial distributions for the doubly degenerate state at $k_x=0$ under the top and side viewpoints, respectively.
  }
  \label{fig6}
\end{figure}

 In this section, we take 1081-SG. 194-4$^2$T13-CA as an example to discuss the PRNL state in its phonon dispersion curves. The results of 128-SG. 194-lon-a with the PRNL state are shown in the Supporting Information (see $\textbf{Figures S65-S66}$). The crystal structure of 1081-SG. 194-4$^2$T13-CA, as exhibited in $\textbf{Figure~\ref{fig5}}$a, belongs the space group $\mathrm{P6}_3 / \mathrm{mmc}$ with number 194. The symmetry operators of 1081-SG. 194-4$^2$T13-CA are generated via $C_{3 z},\left\{C_{2 z} \mid 00 \frac{1}{2}\right\}, C_{2 x y}$, $\mathcal{P}$ and $\mathcal{T}$. Due to our interest in the phonon spectrum, as depicted in $\textbf{Figure~\ref{fig5}}$c, we have disregarded the SOC, resulting in $(\mathcal{P}\mathcal{T})^2=1$.

The enlarged phonon spectrum of 1081-SG. 194-4$^2$T13-CA, displayed in $\textbf{Figure~\ref{fig5}}$d, reveals two crossing points (indicated by black dots) within the frequency ranges of 38.4-39.0 THz. One crossing point is located on the M-K path, while the other is on the K-$\Gamma$ path. Actually, the two points belong to the K-centered closed nodal line on the $k_z=0$ plane since the nodal line is also protected by a glide  mirror symmetry $\left\{M_{z} \mid 00 \frac{1}{2}\right\}$.

\begin{table}
		\caption{Parity information of the 1081-SG. 194-4$^2$T13-CA at the eight TRIM points.}
		\label{Table2}
\begin{tabular*}{\textwidth}{@{\extracolsep{\fill}} ccccccccc}
  \hline\hline
   & \multicolumn{4}{c}{$k_x$=0}  & \multicolumn{4}{c}{$k_x=\pi$} \\
  \cline{2-5}\cline{6-9}

   & $\Gamma$ & $Y$ & $Z$ & $T$ & $X$ & $S$ & $R$ & $U$ \\ \hline
  $n_{+}$ & 22 & 24 & 23 & 23 & 24 & 24 & 23 & 23 \\
  $n_{-}$ & 24 & 22 & 23 & 23 & 22 & 22 & 23 & 23 \\
  $\nu_R$  & \multicolumn{4}{c}{1} & \multicolumn{4}{c}{0}\\
  \hline\hline
\end{tabular*}
\end{table}

Due to $\mathcal{T}$, there must be another closed nodal line around K$'$ on the $k_z=0$ plane. The schematic figure for the $\mathcal{T}$-reversed pair of nodal lines around the K and K$'$ on the $k_z=0$ plane is shown in $\textbf{Figure~\ref{fig5}}$e. To determine the topological character, we examine the $\nu_R$ for vertical 2D planes (i.e., $k_x=0$ and $k_x=\pi$ planes) on the two sides of a K-centered closed nodal line. One finds that the calculated $\nu_R$ for the $k_x=0$ and $k_x=\pi$ planes are 1 and 0, respectively, according to the parity information at four TRIM points and the Wilson loop spectrum for each plane, which are shown in $\textbf{Table~\ref{Table2}}$ and $\textbf{Figure~\ref{fig6}}$b. The switch in topology with $k_x$ from $\nu_R$ = 1 to $\nu_R$ = 0 dictates the existence of PRNL between the two $k_x$ planes. The results of the Wilson loops exhibited in $\textbf{Figure~\ref{fig6}}$a show the nodal line carries a nontrivial second Stiefel-Whitney number (real Chern number) $\nu_R$ = 1 defined on a sphere enclosing a nodal line. In addition, the nodal line features a nontrivial first Stiefel-Whitney number $w$ = 1 defined on a closed path encircling the line based on Equation (2). Therefore, the PRNL states of 1081-SG. 194-4$^2$T13-CA are doubly charged, and the two topological charges~\cite{add8,add13,add46} result in boundary states of different dimensions.

The 1D invariant $w$ of PRNL state leads to the appearance of drumheadlike surface states on the (001) surface of 1081-SG. 194-4$^2$T13-CA. For this surface, the two crossing points on the M-K and K-$\Gamma$ path in the bulk are projected into two surface points on the $\bar{M}$-$\bar{K}$ and $\bar{K}$-$\bar{\Gamma}$ paths in the surface BZ, and a PSS connected to the two surface points is obvious, as shown in $\textbf{Figure~\ref{fig5}}$f. For a better view of the drumheadlike surface state~\cite{add47,add48,add49}, we show the isofrequency surface contour at 38.676 THz of 1081-SG. 194-4$^2$T13-CA on the (001) surface in $\textbf{Figure~\ref{fig5}}$g, where the PSS is highlighted by a black arrow.

The 2D invariant $\nu_R$ of PRNL state leads to the appearance of hinge states on a pair of $\mathcal{P} \mathcal{T}$-related horizontal hinges. As exhibited in $\textbf{Table~\ref{Table2}}$ and $\textbf{Figure~\ref{fig6}}$b, the $k_x=0$ carries a nontrivial $\nu_R$, $k_x=\pi$ carries a trivial $\nu_R$, and a K-centered closed nodal line is located between the two planes. Actually, each slice with a given $k_x$ between the K-centered and K'-centered closed nodal lines is a 2D PRCI with a pair of $\mathcal{P} \mathcal{T}$-related corners. The corners derived from these nontrivial slices between the K-centered and K'-centered closed nodal lines form the hinges for the 3D 1081-SG. 194-4$^2$T13-CA. To verify the PHSs of 1081-SG. 194-4$^2$T13-CA, we take a phononic TB model for the sample of 1081-SG. 194-4$^2$T13-CA with a tube-like geometry, which preserves its $\mathcal{P} \mathcal{T}$ symmetry. Such a tube structure is periodic along x, as shown in $\textbf{Figure~\ref{fig6}}$f. The obtained spectrum plotted in $\textbf{Figure~\ref{fig6}}$c is an enlarged phonon band spectrum (in $k_x$ direction) within the frequency ranges of 38-39.5 THz for a tube sample of 1081-SG. 194-4$^2$T13-CA. One can observe the presence of a phononic hinge band, indicated by a red line, that terminates at the projections of the bulk nodal line. By selecting a doubly degenerate state at $k_x=0$ (see $\textbf{Figure~\ref{fig6}}$d), the spatial distribution for the state is exhibited in $\textbf{Figure~\ref{fig6}}$e and $\textbf{Figure~\ref{fig6}}$f with a different viewpoint. Obviously, the doubly degenerate state appears at one pair of $\mathcal{P} \mathcal{T}$-related hinges of the tube sample of 1081-SG. 194-4$^2$T13-CA.

\subsection{PRTPP state in 52-SG. 141-gis}

\begin{figure}[h]
\centering
  \includegraphics[height=11.5cm]{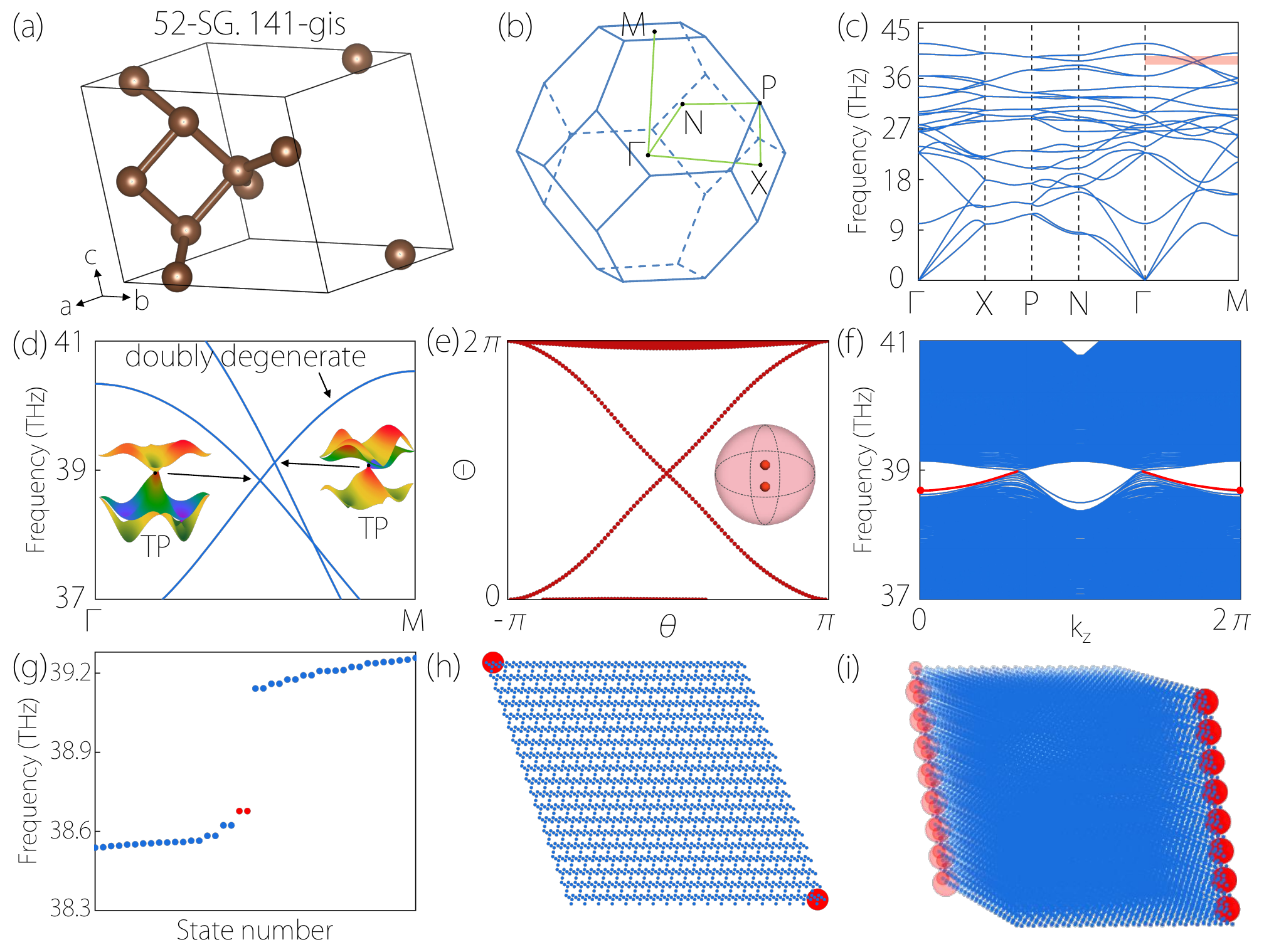}
  \caption{(a) Crystal structure of 52-SG. 141-gis. (b) 3D bulk BZ. (c) Calculated phonon spectrum for 52-SG. 141-gis. (d) Enlarged phonon spectrum within the frequency ranges of 37-41 THz. Insert figure shows the 3D plot of the phononic bands around the PRTPP. (e) The Wilson loop on a sphere enclosing one PRTPP. (f) Phonon spectrum (in $k_z$ direction) within the frequency ranges of 37-41 THz for a sample of 52-SG. 141-gis with the tube-like geometry (as exhibited in (i)). One phononic hinge band is shown in red. (g) Frequency spectrum for a doubly degenerate state (red dot) at $k_z=0$. (h) and (i) The spatial distributions for the doubly degenerate state at $k_z=0$ under the top and side viewpoints, respectively.
  }
  \label{fig7}
\end{figure}

 Analogous to a single Weyl point that can not be annihilated, a single PRNL also exhibits remarkable stability~\cite{add18}. Intriguingly, a PRNL can morph into alternative forms at certain parameters, such as a PRTPP. With additional crystalline symmetry operators, this special form can also stabilize in materials. In this section, we take 52-SG. 141-gis as an example to discuss the PRTPP state in its phonon dispersion curves. The results of the other seven examples are shown in the Supporting Information (see $\textbf{Table S66}$ and $\textbf{Figures S67-S73}$). The crystal structure of 52-SG. 141-gis, as exhibited in $\textbf{Figure~\ref{fig7}}$a, belongs to the space group $14_1 / a m d$ with the number 141.  The calculated phonon spectrum for 52-SG. 141-gis along the symmetry paths $\Gamma$--X--P--N--$\Gamma$--M is shown in  $\textbf{Figure~\ref{fig7}}$c. The phonons belong to bosons, and the SOC is absent, reflecting that the system possesses $\mathcal{P}\mathcal{T}$ symmetry squaring to 1.

The enlarged phonon spectrum of 52-SG. 141-gis is exhibited in $\textbf{Figure~\ref{fig7}}$d. A doubly degenerate quadratic nodal line is formed on the path $\Gamma$-M in the protection of $C_{4 z}$, $M_{y}$, and $\mathcal{P}\mathcal{T}$ symmetry operators, and crosses with two other bands forming a couple of TPs. Two TPs, forming a TPP, are observed at a frequency of 39 THz on the $\Gamma$-M path.

By calculating the $\nu_R$ for horizontal 2D planes (i.e., $k_z=0$ and $k_z=\pi$ planes) on the two sides of TPP based on the parity information at four TRIM points on each plane, one finds that a change in topology with $k_z$ from $\nu_R$ = 1 to $\nu_R$ = 0 dictates the existence of TPP between the two $k_z$ planes (see $\textbf{Table~\ref{Table3}}$). We calculate the Wilson-loop spectrum of a sphere enclosing the TPP and find that it winds once at $\theta=\pi$, thus indicating $\nu_R$ = 1 (see $\textbf{Figure~\ref{fig7}}$e).

\begin{table}
		\caption{Parity information of the 52-SG. 141-gis at the eight TRIM points.}
		\label{Table3}
\begin{tabular*}{\textwidth}{@{\extracolsep{\fill}} ccccccccc}
  \hline\hline
   & \multicolumn{4}{c}{$k_z$=0}  & \multicolumn{4}{c}{$k_z=\pi$} \\
  \cline{2-5}\cline{6-9}

   & $\Gamma$ & $X$ & $S$ & $Y$ & $Z$ & $U$ & $R$ & $Y$ \\ \hline
  $n_{+}$ & 11 & 12 & 11 & 10 & 11 & 10 & 11 & 10 \\
  $n_{-}$ & 11 & 10 & 11 & 12 & 11 & 12 & 11 & 12 \\
  $\nu_R$  & \multicolumn{4}{c}{1} & \multicolumn{4}{c}{0}\\
  \hline\hline
\end{tabular*}
\end{table}

The 2D invariant $\nu_R$ of 52-SG. 141-gis can lead to the PHSs connected to the two pairs of PRTPP, one on the $\Gamma$-M and the other on the $\Gamma$-M' paths (see the schematic figure of PRTPP state in $\textbf{Figure~\ref{fig1}}$c). In order to characterize the topological PHSs originating from the PRTPP of the 52-SG. 141-gis, we utilize a phononic TB model for the 52-SG. 141-gis sample, which is tube-like and has $\mathcal{P} \mathcal{T}$. As depicted in $\textbf{Figure~\ref{fig7}}$i, the constructed 1D tube structure exhibits periodicity along the z-axis. In $\textbf{Figure~\ref{fig7}}$f, we plot an enlarged phonon band spectrum (in $k_z$ direction) of the 1D tube sample with frequencies ranging from 37 to 41 THz. The PHS states, marked by red lines in $\textbf{Figure~\ref{fig7}}$f, are clearly visible as they emerge and connect the projections of the PRTPP. To provide clarity, we have chosen a state at $k_z=0$, indicated by a red dot in $\textbf{Figure~\ref{fig7}}$f. This state has been confirmed to possess double degeneracy, as shown in $\textbf{Figure~\ref{fig7}}$g. The spatial distribution for the doubly degenerate state at $k_z=0$ is given in $\textbf{Figure~\ref{fig7}}$h and $\textbf{Figure~\ref{fig7}}$i, providing confirmation of its location at two $\mathcal{P} \mathcal{T}$-related hinges.

\subsection{PRDP state in 132-SG. 191-3,4T157}

\begin{figure}[h]
\centering
  \includegraphics[height=12cm]{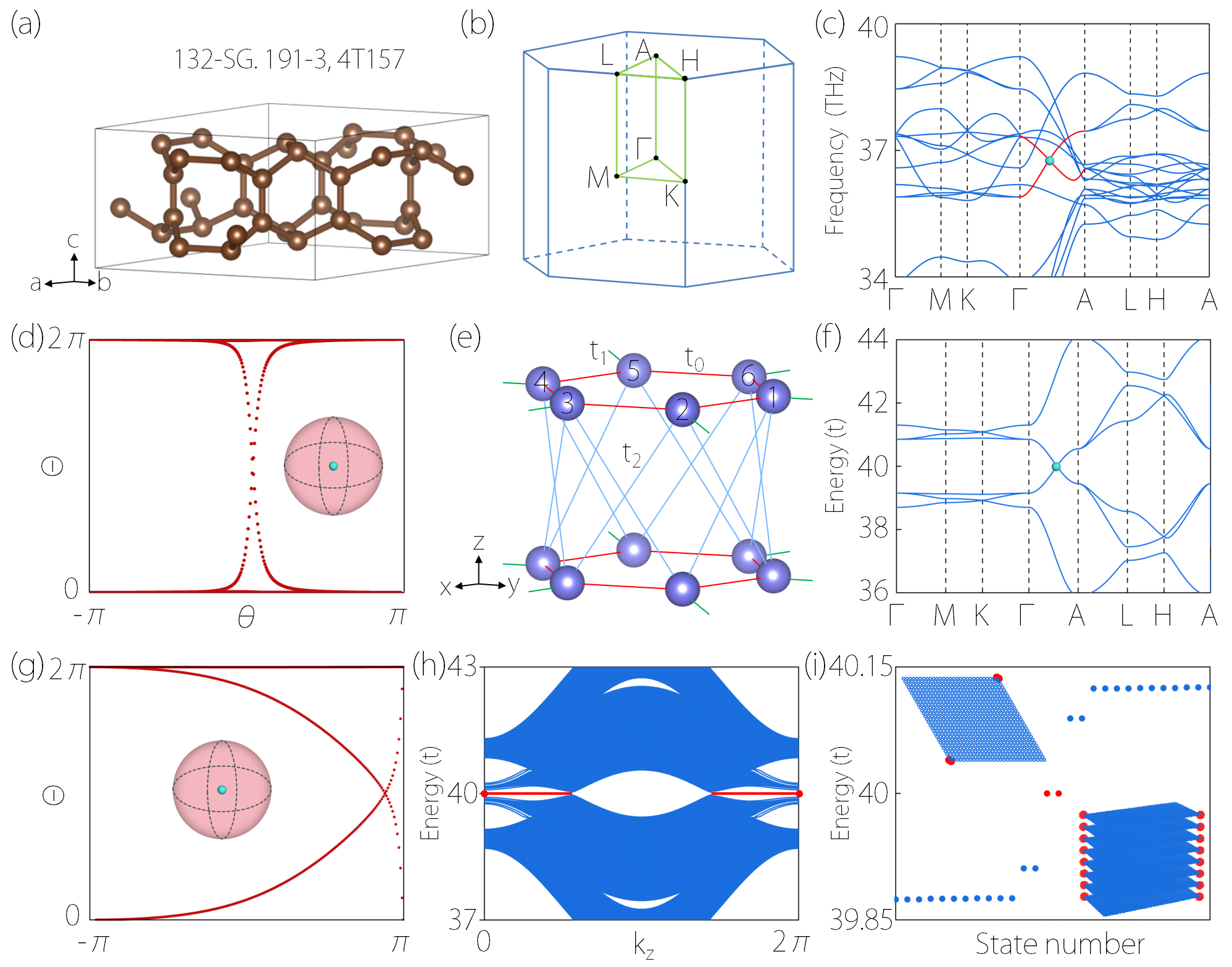}
  \caption{(a) Crystal structure of 132-SG. 191-3,4T157. (b) 3D bulk BZ. (c) Enlarged phonon spectrum within the frequency ranges of 34-40 THz. A DP (marked by a green ball) appears on the $\Gamma$-A path. (d) The Wilson loop on a sphere enclosing the DP of 132-SG. 191-3,4T157. (f) 3D TB model with t$_1$ = t, t$_0$ = 0.8t, t$_2$ = 0.35t, and onsite energy of 40t. (f) Bulk band structure. A DP (marked by a green ball) appears on the $\Gamma$-A path. (g) The Wilson loop on a sphere enclosing the DP of the TB model. (h) The spectrum of a 1D tube geometry (in $k_z$ direction). (i) Energy spectrum for a doubly degenerate state (red dot) at $k_z=0$. Inserts figures are the spatial distributions for the state at $k_z=0$ under the top and side viewpoints.
  }
  \label{fig8}
\end{figure}

The PRNL can also shrink into a PRDP and stabilize in materials with specific crystalline symmetries. The presence of a 3D PRDP state was identified in 10 out of 1192 carbon allotropes. As an example, the enlarged phonon spectrum within the frequency ranges of 34-40 THz of 132-SG. 191-3,4T157 is exhibited in $\textbf{Figure~\ref{fig8}}$c. The results of other candidates are shown in the Supporting Information (see $\textbf{Table S67}$ and $\textbf{Figures S74-S83}$). The 132-SG. 191-3,4T157 with a P6/mmm space group (number 191) and hosts the $\mathcal{P} \mathcal{T}$ symmetry. As shown in $\textbf{Figure~\ref{fig8}}$c, one accidental DP appears on the $\Gamma$-A path. Note that quadratic nodal lines are formed along the high symmetric path $\Gamma$-A in the protection of $C_{6 z}$ and $\mathcal{P} \mathcal{T}$ symmetry operators. The quadratic nodal lines with different $C_{6 z}$ eigenvalues intersect, forming DPs. To indicate the real topology for the DP on the $\Gamma$-A path, we calculate the Wilson loop on a sphere enclosing one DP, which is nontrivial with $\nu_R$ = 1 (see $\textbf{Figure~\ref{fig8}}$d).

Regrettably, the phononic bands intersecting that form the PRDP on the $\Gamma$-A path are not clear, and there are numerous neighboring phononic bands visible. Identifying the PHSs in such a setting is challenging because of the perturbation caused by other phononic bands. Nevertheless, in the following, we present a six-band TB model for the P6/mmm space group (number 191) (see $\textbf{Figure~\ref{fig8}}$e and the Hamiltonian in Supporting Information) to show the appearance of $\mathcal{P} \mathcal{T}$-related hinges.

$\textbf{Figure~\ref{fig8}}$f plots the bulk band structure that exhibits one DP on the $\Gamma$-A path. Moreover, the DP hosts a nontrivial $\nu_R$, as demonstrated by the calculated Wilson loop spectrum in $\textbf{Figure~\ref{fig8}}$g. Then, we construct a tube geometry (see the insert figure of $\textbf{Figure~\ref{fig8}}$i) and exhibit the tube spectrum, in which the hinge modes are marked with red color (see $\textbf{Figure~\ref{fig8}}$h). We also select a state at $k_z=0$ (see the red dot in $\textbf{Figure~\ref{fig8}}$h), and demonstrate such a state is distributed at the $\mathcal{P} \mathcal{T}$-related hinges of the system (see $\textbf{Figure~\ref{fig8}}$i).

\section{Conclusion}

In summary, based on the first-principles calculations and symmetry analysis, we propose that the 3D carbon allotropes are a valuable framework for exploring the real topological phononic states and related PHSs. Out of a total of 1192 3D carbon allotropes, 65 of them have a PRCI state, 2 of them have a PRNL state, 10 of them have a PRDP state, and 8 of them have a PRTPP state, respectively. The PRCI, PRNL, PRDP, and PRTTP states are protected by $\mathcal{P} \mathcal{T}$ and characterized by a nontrivial $\nu_R$. The real topology of PRCI, PRNL, PRDP, and PRTTP states in 3D carbon allotropes is manifested in  the topological PHSs. We suggest 27-SG. 166-pcu-h, 1081-SG. 194-4$^2$T13-CA, 52-SG. 141-gis, and 132-SG. 191-3,4T157 as ideal examples hosting clean topological PHSs. The PRNLs of 52-SG. 141-gis also exhibit a nontrivial $w$ in addition to the nontrivial $\nu_R$. This suggests the possibility of attaining multi-dimensional boundaries (PSSs and PHSs) within a single target with PRNL states, as opposed to a target with conventional phononic NL states. Our research expands on the scope of topological phononic materials by providing a material realization that includes four distinct forms of real topological phononic states. Additionally, we elucidate the essential characteristics of the topological boundary modes associated with these real topological phonons.



\medskip

%

\begin{thebibliography}{}
\bibitem{add1} J. Wang, S. C. Zhang, \textit{Nat. Mater.}, \textbf{2017}, 16, 1062.
\bibitem{add2} C.-K. Chiu, J. C. Y. Teo, A. P. Schnyder, S. Ryu, \textit{Rev. Mod. Phys.} \textbf{2016}, 88, 035005.
\bibitem{add3} X.-L. Qi, S.-C. Zhang, \textit{Rev. Mod. Phys.} \textbf{2011}, 83, 1057.
\bibitem{add4} B. Singh, H. Lin, A. Bansil, \textit{Adv. Mater.}, \textbf{2023}, 35, 2201058.
\bibitem{add5} T. Zhang, Y. Jiang, Z. Song, H. Huang, Y. He, Z. Fang, H.  Weng, C. Fang,  \textit{Nature}, \textbf{2019}, 566, 475.
\bibitem{add13} J. Ahn, S. Park, D. Kim, Y. Kim, B. J. Yang, \textit{Chin. Phys. B}, \textbf{ 2019}, 28, 117101.
\bibitem{add7} Y. X. Zhao and Y. Lu, \textit{Phys. Rev. Lett.} \textbf{2017}, 118, 056401.
\bibitem{add8} J. Ahn, D. Kim, Y. Kim, B.-J. Yang, \textit{Phys. Rev. Lett.} \textbf{2018}, 121, 106403.
\bibitem{add19} P. M. Lenggenhager, X. Liu, T. Neupert, T. Bzdu\v{s}ek, \textit{Phys. Rev. B} \textbf{2022},106, 085129.
\bibitem{add19a} C. Fang, Y. Chen, H.-Y. Kee, L. Fu, \textit{Phys. Rev. B} \textbf{2015}, 92, 081201(R).
\bibitem{add7a} M. G. Vergniory, L. Elcoro, C. Felser, N. Regnault, B. A. Bernevig, Z. Wang, \textit{Nature} \textbf{2019}, 566, 480.
\bibitem{add7b} T. Zhang, Y. Jiang, Z. Song, H.  Huang, Y. He, Z. Fang, H. Weng, C. Fang, \textit{Nature} \textbf{2019}, 566, 475.
\bibitem{add7c} Y. Xu, L. Elcoro, Z. D. Song, B. J. Wieder, M. G. Vergniory, N. Regnault, Y. Chen, C. Felser, B. A. Bernevig, \textit{Nature} \textbf{2020}, 586, 702.
\bibitem{add7d} B. Bradlyn, L. Elcoro, J. Cano, M. G. Vergniory, Z. Wang, C. Felser, M. I. Aroyo, B. A. Bernevig, \textit{Nature} \textbf{2017}, 547, 298.
\bibitem{add7e} R. Chen, H. C. Po, J. B. Neaton, A. Vishwanath, \textit{Nat. Phys.} \textbf{2018}, 14, 55.
\bibitem{add7ee}B. A. Bernevig, C. Felser, H. Beidenkopf, \textit{Nature} \textbf{2022}, 603, 41.
\bibitem{add7ef} B. J. Wieder, B.  Bradlyn, J. Cano, Z. Wang, M. G. Vergniory, L. A. A. Soluyanov, C. Felser, T. Neupert, N. Regnault, B. A. Bernevig, \textit{Nat. Rev. Mat.} \textbf{2022}, 7, 196.
\bibitem{add7eg} F. Tang, H. C. Po, A. Vishwanath, X. Wan, \textit{Nat. Phys.} \textbf{2019}, 15, 470.
\bibitem{add7eh}J. Xiao, B. Yan, \textit{Nat. Rev. Mat.} \textbf{2021}, 3, 283.
\bibitem{add7f} Y. Xu, M. G. Vergniory, D.-S. Ma, J. L. Ma$\tilde{\rm{n}}$es, Z.-D. Song, B. A. Bernevig, N. Regnault, L. Elcoro, \textit{Science} \textbf{2024}, 384, eadf8458.
\bibitem{add16} X. T. Zeng, B. B.Liu, F. Yang, Z. Zhang, Y. X. Zhao, X. L. Sheng, S. A. Yang, S. A. \textit{Phys. Rev. B} \textbf{2023}, 108, 075159.
\bibitem{add12} C. Chen, X.-T. Zeng, Z. Chen, Y. X. Zhao, X.-L. Sheng, S. A. Yang, \textit{Phys. Rev. Lett.} \textbf{2022}, 128, 026405.
\bibitem{add14}H. Xue, Z. Chen, Z. Cheng, J. Dai, Y. Long, Y. Zhao, B. Zhang, \textit{Nat. Commun.} \textbf{2023}, 14, 4563.
\bibitem{add15}Y. Pan, C. Cui, Q. Chen, F. Chen, L. Zhang, Y. Ren, N. Han, W. Li, X. Li, Z.-M. Yu, H. Chen, Y. Yang, \textit{Nat. Commun.} \textbf{2023}, 14, 6636.
\bibitem{add15a}X. Xiang, X. Ni, F. Gao, X. Wu, Z. Chen, Y.-G. Peng,  X.-F. Zhu, \textit{arXiv preprint} \textbf{2023} arXiv: 2304.12735.
\bibitem{add17} J. Wang, T.-T. Zhang, Q. Zhang, X. Cheng, W. Wang, S. Qian, Z. Cheng, G. Zhang, X. Wang, \textit{Adv. Funct. Mater.} \textbf{2024}, 2316079.
\bibitem{add18} S. Qian, Y. Li, C. C. Liu, \textit{Phys. Rev. B} \textbf{2023}, 108, L241406.
\bibitem{add20} Y. Wang, C. Cui, R.-W. Zhang, X. Wang, Z.-M. Yu, G.-B. Liu, Y. Yao, \textit{Phys. Rev. B} \textbf{2024}, 109, 195101.
\bibitem{add21}Y. Li, S. Qian, C. C. Liu, \textit{arXiv preprint} \textbf{2023} arXiv: 2309.01566.
\bibitem{add21b}L. Luo, H.-X. Wang, Z.-K. Lin, B. Jiang, Y. Wu, F. Li, J.-H. Jiang, \textit{Nat. Mater.} \textbf{2021}, 20, 794.
\bibitem{add10} X.-L. Sheng, C. Chen, H. Liu, Z. Chen, Z.-M. Yu, Y. X. Zhao, S. A. Yang, \textit{Phys. Rev. Lett.} \textbf{2019}, 123, 256402.
\bibitem{add22}X. Zhang, T. He, Y. Liu, X. Dai, G. Liu, C. Chen, W. Wu, J. Zhu, S. A. Yang, \textit{Nano Lett.} \textbf{2023}, 23, 7358.
\bibitem{add23} X. Wang, X.-P. Li, J. Li, C. Xie, J. Wang, H. Yuan, W. Wang, Z. Cheng, Z.-M. Yu, G. Zhang, \textit{Adv. Funct. Mater.} \textbf{2023}, 33, 2304499.
\bibitem{add24}G. Liu, H. Jiang, Z. Guo, X. Zhang, L. Jin, C. Liu, Y. Liu, \textit{Adv. Sci.} \textbf{2023}, 10, 2301952.
\bibitem{add25} J. Gong, Y. Wang, Y. Han, Z. Cheng, X. Wang, Z.-M. Yu, Y. Yao, \textit{Adv. Mater.} \textbf{2024}, doi: 10.1002/adma. 202402232.
\bibitem{add26} M. Pan, D. Li, J. Fan, H. Huang, \textit{npj Comput. Mater.} \textbf{2022}, 8, 1.
\bibitem{add27} Y. Liu, X. Chen, Y. Xu, \textit{Adv. Funct. Mater.} \textbf{2020}, 30, 1904784.
\bibitem{add28} Y. Liu, N. Zou, S. Zhao, X. Chen, Y.  Xu, W. Duan, \textit{Nano Lett.} \textbf{2022}, 22, 2120-2126.
\bibitem{add29}Z. K. Ding, Y. J. Zeng, W. Liu, L. M. Tang, K. Q. Chen, \textit{Adv. Funct. Mater.} \textbf{2024}, 2401684.
\bibitem{add30}T. Yang, J. Wang, X. P. Li, X. Wang, Z. Cheng, W. Wang, G. Zhang, \textit{Matter} \textbf{2024}, 7, 320-350.
\bibitem{add31} J. Li, J. Liu, S. A. Baronett, M. Liu, L. Wang, R. Li, Y. Chen, D. Li, Q. Zhu, X.-Q. Chen, \textit{Nat. Commun.} \textbf{2021}, 12, 1204.
\bibitem{add32} B. Peng, Y. Hu, S. Murakami, T. Zhang, B. Monserrat, \textit{Sci. Adv.} \textbf{2020}, 6, eabd1618.
\bibitem{add33} Y. Xu, M. G. Vergniory, D. S. Ma, J. L. Manes, Z. D. Song, B. A. Bernevig, N. Regnault, L. Elcoro \textit{arXiv preprint} \textbf{2022} arXiv: 2211.11776.
\bibitem{add34} X. Wang, T. Yang, Z. Cheng, G. Surucu, J. Wang, F. Zhou, Z. Zhang, G. Zhang, \textit{Appl. Phys. Rev.} \textbf{2022}, 9, 041304.
\bibitem{add35} J. Li, J. Li, J. Tang, Z. Tao, S. Xue, J. Liu, H. Peng, X.-Q. Chen, J. Guo, X. Zhu, \textit{Phys. Rev. Lett.} \textbf{2023}, 131, 116602.
\bibitem{add36} J. Li, Y. Liu, J. Bai, C. Xie, H. Yuan, Z. Cheng, W. Wang, X. Wang, G. Zhang, \textit{Appl. Phys. Rev.} \textbf{2023}, 10, 031416.
\bibitem{add37} H. Mu, B. Liu, T. Hu, Z. Wang, \textit{Nano Lett.} \textbf{2022}, 22, 1122.
\bibitem{add38} J. Zhu, W. Wu, J. Zhao, C. Chen, Q. Wang, X.-L. Sheng, L. Zhang, Y. X. Zhao, S. A. Yang, \textit{Phys. Rev. B} \textbf{2022}, 105, 085123.
\bibitem{add39} M. Pan, H. Huang, \textit{Phys. Rev. B} \textbf{2022}, 106, L201406.
\bibitem{add21a} R. Hoffmann, A. A. Kabanov, A. A. Golov, D. M. Proserpio, \textit{Angew. Chem. Int. Ed.}, \textbf{2016}, 55, 10962-10976.
\bibitem{add40} https://www.sacada.info/ (accessed: January 2024).
\bibitem{add41} Y. Kopelevich, P. Esquinazi, \textit{Adv. Mater.} \textbf{2007}, 19, 4559-4563.
\bibitem{add42} B. A. Fairchild, S. Rubanov, D. W. M. Lau, M. Robinson, I. Suarez-Martinez, N. Marks, A. D. Greentree, D. McCulloch, S. Prawer, \textit{Adv. Mater.} \textbf{2012}, 24.
\bibitem{add43} W. Zhou, X. Bai, E. Wang, S. Xie, \textit{Adv. Mater.} \textbf{2009}, 21, 4565-4583.
\bibitem{add44} N. O. Weiss, H. Zhou, L. Liao, Y. Liu, S. Jiang, Y. Huang, X. Duan, \textit{Adv. Mater.} \textbf{2012}, 24, 5782-5825.
\bibitem{add45} M. Chen, R. Guan, S. Yang, \textit{Adv. Sci.} \textbf{2019}, 6, 1800941.
\bibitem{add46} K. Shiozaki, M. Sato, K. Gomi, \textit{Phys. Rev. B} \textbf{2017}, 95, 235425.
\bibitem{add47} I. Belopolski, K. Manna, D. S. Sanchez, G. Chang, B. Ernst, J. Yin, S. S. Zhang, T. Cochran, N. Shumiya, H. Zheng, B. Singh, G. Bian, D. Multer, M. Litskevich, X. Zhou, S.-M. Huang, B. Wang, T.-R. Chang, S.-Y. Xu, A. Bansil, C. Felser, H. Lin, M. Z. Hasan, \textit{Science}, \textbf{2019}, 365, 1278-1281.
\bibitem{add48} Q. Yan, R. Liu, Z. Yan, B. Liu, H. Chen, Z. Wang, L. Lu, \textit{Nat. Phys.} \textbf{2018}, 14, 461-464.
\bibitem{add49} W. Deng, J. Lu, F. Li, X. Huang, M. Yan, J. Ma, Z. Liu, \textit{Nat. Commun.} \textbf{2019}, 10, 1769.













\end{thebibliography}

\clearpage
\begin{figure}
\textbf{Table of Contents}\\
\medskip
  \includegraphics[height=7.8cm]{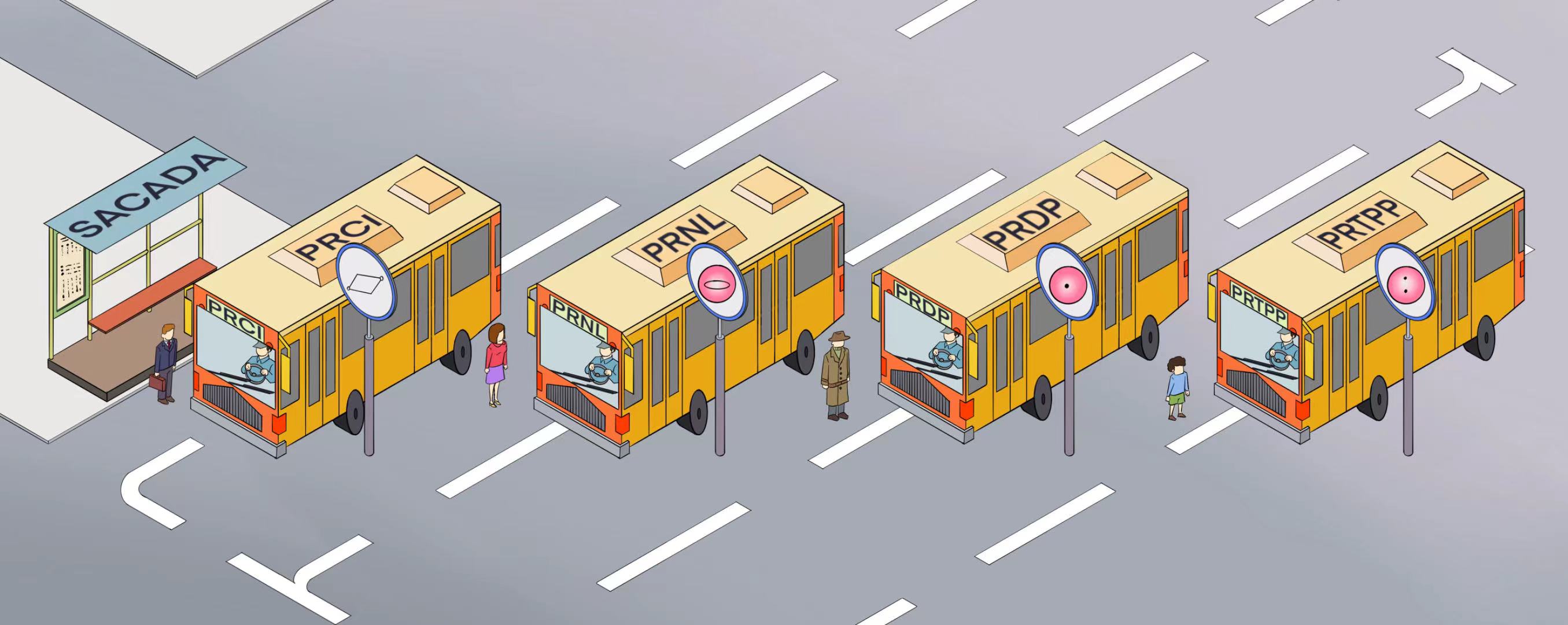}
  \medskip
  \caption*{ Our findings indicate that among the 1192 3D carbon allotropes in the Samara Carbon Allotrope Database (SACADA), 65 possess a phononic real Chern insulating (PRCI) state, 2 possess a phononic real nodal line (PRNL) state, 10 possess a phononic real Dirac point (PRDP) state, and 8 possess a phononic real triple-point pair (PRTPP), respectively. The real topology of PRCI, PRNL, PRDP, and PRTTP states in 3D carbon allotropes is manifested in the second-order phononic hinge states.}
\end{figure}

\end{document}